\documentstyle[psfig,prb,aps,multicol]{revtex}
\begin{document}
\title{Diamagnetic response of a normal-metal--superconductor 
  proximity system at arbitrary impurity concentration} 
\author{W.\ Belzig$^a$, C.\ Bruder$^a$, and A.~L.\ Fauch\`ere$^b$} 
\address{
 $^a$Institut f{\"u}r Theoretische Festk{\"o}rperphysik, Universit{\"a}t
 Karlsruhe, D-76128 Karlsruhe, Germany\\
 $^b$Theoretische Physik, Eidgen\"ossische Technische Hochschule,
 CH-8093 Z{\"u}rich, Switzerland
}
\date{\today}
\maketitle
\begin{abstract}
  We investigate the magnetic response of normal-metal--superconductor
  proximity systems for arbitrary concentrations of impurities and at
  arbitrary temperatures. Using the quasiclassical theory of superconductivity
  a general
  linear-response formula is derived which yields a non-local
  current-field relation in terms of the zero-field Green's functions.
  Various regimes between clean-limit and dirty-limit response
  are investigated by analytical methods and by solving the general
  formula numerically. In the ballistic regime, a finite
  mean free path reduces the non-locality and leads to a stronger
  screening than in the clean limit even for a mean free path much
  larger than the system size. Additionally, the range of the kernel
  describing the non-locality is strongly temperature-dependent in
  this case. In the diffusive limit we find a cross-over from local to
  non-local screening, which restricts the applicability of the
  dirty-limit theory.
\end{abstract}
%74.50.+r Proximity effects, weak links, tunneling phenomena, and Josephson
% effects
%74.25.Ha Magnetic properties
%73.23.-b Mesoscopic systems

\begin{multicols}{2}

\section{Introduction}
A normal metal in good metallic contact to a superconductor acquires
induced superconducting properties. The basic features of this {\it
  proximity effect} were already well understood in the sixties. One
of these properties is the diamagnetic screening of an applied
magnetic field, which has been studied in a series of
experimental\cite{mota:82,pobell:87,mota:90,mota:94} and theoretical
works\cite{degennes,zaikin,higashitani,belzig,fauchere}.

Already early in this development it was recognized that the relevant
length scale governing the superconducting correlations is given by
the thermal and impurity dependent coherence lengths in the normal
metal. The thermal length is given by $\xi_T=v_{\text{F}}/2\pi T$ in
the clean limit (mean free path $l\to\infty$) and $\xi_D=(\xi_T
l/3)^{1/2}$ in the dirty limit ($l\to 0$). The finite thickness $d$ of
the normal-metal layer is an intrinsic geometric length scale of the
proximity effect. The interplay between the three length scales
$\xi_T$, $l$, and $d$ is relevant for the behavior of the microscopic
quantities such as the spatial decay of the pair amplitude or the
spectral density of states. In this paper we investigate the range
between the clean and the dirty limiting cases and find several
intermediate regimes of interest, differing by the relative magnitudes
of $\xi_T$, $l$, and $d$.

The theory of linear diamagnetic response of clean and dirty N-S
proximity systems has already been studied
extensively\cite{degennes,zaikin,belzig}. While the dirty-limit theory
was found to be in agreement with early experimental work
\cite{oda:80}, the samples studied in more recent
experiments\cite{mota:89,oda:95} fail to be described satisfactorily
by either the clean or the dirty limit. They exhibit the qualitatively
different behavior of an intermediate regime. In a previous
paper\cite{belzig} we were able to fit an experiment in the
low-temperature regime with the dirty-limit theory. However, the data
for temperatures $T> v_{\text{F}}/d$ could not be reproduced by the
dirty limit theory. These experiments also show clear deviations from
the clean-limit theory, even if the finite transparency of the
interface is accounted for\cite{higashitani,fauchere}. On the other
hand, the breakdown field seen in the nonlinear response to a magnetic
field of some of the same samples was found to agree fairly well with
the clean-limit theory\cite{fauchere}. In this paper, guided by a
numerical study, we classify the intermediate regimes and show how the
qualitative discrepancies between theory and experiment can be
resolved. In recent experiments on relatively clean samples a
low-temperature anomaly was reported\cite{mota:89,mota:90,mota:94}.
The nearly perfect screening at $T\approx v_{\text{F}}/d$ was found to
be reduced as the temperature was lowered further. This re-entrance
effect has not been explained up to now. We will not address this
problem directly here, but rather provide an understanding of those
facets of the proximity effect which are the necessary basis for
further investigations.

From a theoretical point of view there is a major qualitative
difference between the magnetic response in the dirty versus the clean
limit. In the dirty limit the current-field relation is local and the
screening can be almost complete. This is in strong contrast to the
clean limit, where the current-field relation is completely nonlocal
and the current depends on the vector potential integrated over the
whole normal-metal layer \cite{zaikin}. As a consequence, there is an
overscreening effect, i.~e., the magnetic field reverses its sign
inside the normal metal, and the magnetic susceptibility is limited to
$3/4$ of that of a perfect diamagnet. In the intermediate regime, the
superfluid density and the range of the current-field relations, both
diminishing with decreasing mean free path, are shown to affect the
screening ability in a contrary way and thus compete in the magnetic
response.

We investigate the magnetic response at arbitrary impurity
concentrations starting from the quasi-classical Green's functions in
absence of the fields, which are discussed in Section
\ref{sec:proxieffect}. On this basis, we develop a theory of linear
current response in Section \ref{sec:linmagresponse}. We produce a
general result with the well-known structure: the current density
$j(x)$ is given by a convolution of a kernel $K(x,x')$ with the vector
potential $A(x)$. The kernel is given explicitly in terms of the
Green's functions in the absence of the fields. Our formula easily
yields the basic constitutive relations of the London\cite{london}-
and Pippard\cite{pippard}-type for a superconductor. With the help of
this kernel we calculate the magnetic susceptibility of a proximity
system at arbitrary impurity concentrations in
Section~\ref{sec:magresponse}. We find that the impurities have
non-trivial consequences on the magnetic response. The range of the
integral kernel can be strongly temperature dependent, and can be
given by either $\xi_T$ or $l$. In particular, we show that even for
$l$ considerably larger that $d$ the spatial dependence of the
integral kernel strongly enhances the magnetic response, as compared
to the clean limit.

\section{Quasiclassical Equations and Proximity Effect}
\label{sec:proxieffect}
The basic set of equations appropriate for describing spatially
inhomogeneous superconductors was developed by
Eilenberger\cite{eilenberger} and by Larkin and Ovchinnikov
\cite{larkin} (for a recent collection of papers on the quasiclassical
method, see \cite{rainersauls}). They are transport-like equations for
the quasiclassical Green's functions, i.e., the energy-integrated
Gorkov Green's functions, that are derived from the Gorkov equations
under the assumption that the length scales relevant for
superconductivity are much larger than atomic length scales. We treat
the presence of elastic impurities within the Born approximation (the
full T-matrix-formalism has been shown to lead to quantitative
changes\cite{herath}). The Eilenberger equations take the form
($e=|e|$)
\begin{eqnarray}
 \label{eilenberger}
 \lefteqn{-\bbox{v}_{\text{F}}(\bbox{\nabla}+2ie\bbox{A}(\bbox{x}))
 f(\bbox{v}_{\text{F}},\bbox{x}) =}\\ & & 
 (2\omega+\frac{1}{\tau} \langle g(\bbox{x})\rangle)
 f(\bbox{v}_{\text{F}},\bbox{x})- (2\Delta(\bbox{x})+\frac{1}{\tau} \langle
 f(\bbox{x})\rangle) g(\bbox{v}_{\text{F}},\bbox{x})\nonumber \\\nonumber
 \lefteqn{ \bbox{v}_{F}(\bbox{\nabla}-2ie\bbox{A})
 f^{\dagger}(\bbox{v}_{\text{F}},\bbox{x}) = }\\
 & & (2\omega+\frac{1}{\tau} \langle
 g(\bbox{x})\rangle)f^{\dagger}(\bbox{v}_{\text{F}},\bbox{x})-
 (2\Delta(\bbox{x})+\frac{1}{\tau} \langle f^{\dagger}(\bbox{x})\rangle)
 g(\bbox{v}_{\text{F}},\bbox{x})\nonumber \\\nonumber
 \lefteqn{-\bbox{v}_{F}\bbox{\nabla}g(\bbox{v}_{\text{F}},\bbox{x})=}
 \\\nonumber & &(\Delta(\bbox{x})+\frac{1}{2\tau}\langle f(\bbox{x})\rangle)
 (f(\bbox{v}_{\text{F}},\bbox{x})-f^{\dagger}(\bbox{v}_{\text{F}},\bbox{x})) 
\end{eqnarray}
These are three coupled differential equations for the normal
(diagonal) Green's function $g$ and the anomalous (off-diagonal)
Green's functions $f$ and $f^{\dagger}$. They depend on Matsubara
frequency $\omega=\pi T(2n+1)$, the elastic scattering time
$\tau=l/v_{\text{F}}$, and the Fermi velocity $\bbox{v}_{\text{F}}$,
\mbox{$\langle \dots\rangle $} denoting the average over the Fermi
surface ($\hbar=c=1$ throughout).  The superconducting order parameter
$\Delta$ is taken to be real.  We note that the $\omega$-dependence of
the Green's functions has been omitted in our notation. The Green's
functions obey the normalization condition
\begin{equation}
 \label{norm}
 g^{2}(\bbox{v}_{\text{F}},\bbox{x}) +
 f^{}(\bbox{v}_{\text{F}},\bbox{x})
 f^{\dagger}(\bbox{v}_{\text{F}},\bbox{x})=1
\end{equation}
and fulfill the symmetry relations
$g^{*}(-\bbox{v}_{F},\bbox{x}) = g(\bbox{v}_{F},\bbox{x})$ and
$f^{*}(-\bbox{v}_{F},\bbox{x}) = f^{\dagger}(\bbox{v}_{F},\bbox{x})$ .
The current is given by
\begin{equation}
 \label{current}
 \bbox{j}(\bbox{x})=\frac{i2ep_{F}m}{\pi}T\sum_{\omega>0}
 \langle \bbox{v}_{F}g(\bbox{v}_{F},\bbox{x})\rangle \; ,
\end{equation}
and depends only on the imaginary part of $g$, due to the above symmetry
relations.

\begin{figure}[htbp]
 \begin{center}
 \leavevmode
 \psfig{figure=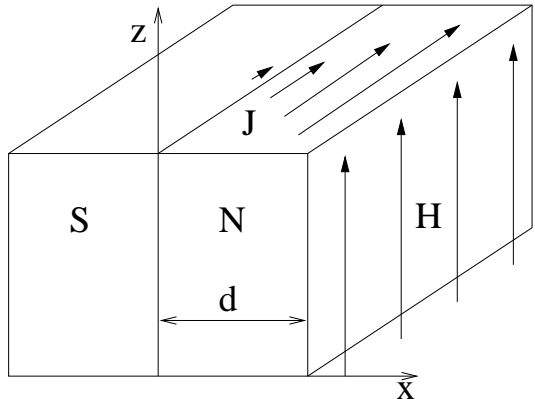,width=7cm}
\narrowtext
 \caption{Geometry of the proximity model system. The thickness
   of the superconductor is assumed to be much greater than $\xi_0$,
   the pair potential is taken real and assumed to follow a step
   function: $\Delta(x)=\Delta\Theta(-x)$. In our gauge, the screening
   current and the vector potential are parallel to the NS interface.
   The interface is assumed to be perfect and the normal-metal--vacuum
   boundary to be specularly reflecting.  }
 \label{fig:geometry}
 \end{center}
\end{figure}

In this paper, we consider a system shown in Fig.~\ref{fig:geometry}
consisting of a normal-metal layer of thickness $d$, which is in ideal
contact with a semi-infinite superconductor.  A magnetic field
$(0,0,H)$ is applied parallel to the metal surface, producing
screening currents $(0,j(x),0)$ along the surface, which depend on the
coordinate $x$. The pair potential is taken to be a step function
$\Delta(x)=\Delta\Theta(-x)$. Assuming a thickness $d \gg
\xi_0=v_{\text{F}}/2\pi T_c$, we can neglect the self-consistency of
the pair potential. Furthermore, we assume specular reflection at the
normal-metal--vacuum boundary.

In the absence of external fields (we denote the corresponding Green's
functions by $g_0$, $f_0$ and $f_0^{\dag}$) Eqs.~(\ref{eilenberger})
reduce to
\begin{eqnarray}
 \label{eilenreal}
 -{v}_{x}\frac{d}{dx}f_{0}(v_{\text{x}},x)&=&
 2\tilde{\omega}(x)f_{0}(v_{\text{x}},x) -
 2\tilde{\Delta}(x)g_{0}(v_{\text{x}},x) 
 \\\nonumber
 {v}_{x}\frac{d}{dx}f^{\dagger}_{0}(v_{\text{x}},x)&=&
 2\tilde{\omega}(x)f_{0}^{\dagger}(v_{\text{x}},x) -
 2\tilde{\Delta}(x)g_{0}(v_{\text{x}},x)\; .
\end{eqnarray}
We have introduced the effective frequency $\tilde{\omega}(x) = \omega
+ \langle g_{0}(x)\rangle /2\tau$ and pair potential
$\tilde{\Delta}(x)=\Delta(x)+\langle f_{0}(x)\rangle /2\tau$.
Equations (\ref{eilenreal}) imply that
$f_{0}(v_{\text{x}},x)=f_{0}^{\dagger}(-v_{\text{x}},x)$ and, since
$\langle f_{0}(x)\rangle =\langle f_{0}^{\dag}(x)\rangle ^{*}$, for
real $\Delta$ we obtain a real $\langle f_{0}(x)\rangle$, too.

\begin{figure}[htbp]
 \begin{center}
\psfig{figure=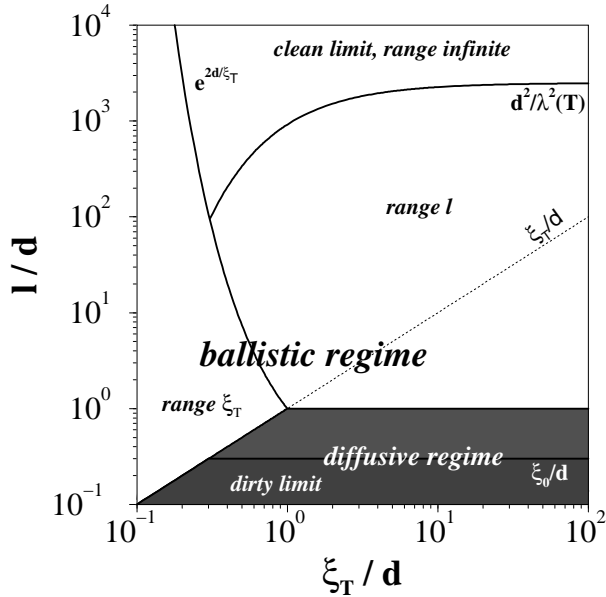,width=8cm} 
\caption{Dependence of the magnetic response on thermal length 
  $\xi_T=v_{\text{F}}/2\pi T$ and mean free path $l$. 
  In the ballistic regime $l\gg {\rm min}\{\xi_T,d\}$ 
  we distinguish three regions: (a) the clean limit with infinite 
  range of the kernel exhibiting a reduced diamagnetism (overscreening),
  (b) the quasi-ballistic limit with finite range $\xi_T$ increasing screening 
  at large temperatures, (c) the ballistic limit where the finite range 
  $l$ enhances the screening although $l\gg d$. 
  In the diffusive regime $l\ll \xi_T,d$ the range of the kernel is given 
  by $l$. The dirty limit with nearly isotropic Green's functions is 
  restricted to $l\ll \xi_0, \xi_T, d$. Note that the current-field 
  relations can still be local or non-local depending on the relative
  size of penetration depth and mean free path. For comparison, the 
  conventionally assumed border line between clean and dirty limits 
  ($l=\xi_T$) is indicated by a dotted line.}
 \label{fig:regime}
 \end{center}
\end{figure}

Depending on the relative size of the thermal length $\xi_T$, the mean
free path $l$, and the thickness $d$ we distinguish the ballistic, the
dirty and the intermediate diffusive regime that are discussed in the
following subsections. These regimes are also shown in
Fig.~\ref{fig:regime}.

\subsection{Ballistic regime}
The ballistic regime is limited by $l\gg {\rm min}\{\xi_T,d\}$, 
which ensures a ballistic
propagation of the electrons over the thickness or the thermal length
of the normal layer, respectively.
As a limiting case, for $l\rightarrow \infty$ (clean limit)
Eq.~(\ref{eilenreal}) may be solved analytically. For $T_c\gg
T_A\equiv v_{\text{F}}/2\pi d$, the solution in the normal metal takes
the form \cite{zaikin}
\begin{eqnarray}
 \label{cleanreal}
 f_0(v_{\text{x}},x) & = & e^{\frac{\displaystyle
 2\omega}{\displaystyle v_{\text{x}}} {\displaystyle (d-x)}}/\cosh\chi_{d}
 \\\nonumber
 f_0^{\dagger}(v_{\text{x}},x) &=& f(-v_{\text{x}},x) , \quad 
\chi_{d}=\frac{2\omega d}{|v_{\text{x}}|}\; .
\end{eqnarray}
At temperatures above the Andreev temperature, $T \gg T_A$, only the
first Matsubara frequency $\omega=\pi T$ is relevant and the decay of
the $f$-function is governed by $\xi_T=v_{\text{F}}/2\pi T$. $T_A$
determines the temperature at which the $f$-function acquires a finite
value at the outer boundary.

An estimation using the Eilenberger equation Eq.~(\ref{eilenreal})
easily shows that the clean-limit solution is valid for
\begin{eqnarray} 
\label{eq:cleanvalid1}
l\gg d \exp(2d/\xi_T) \quad & \text{if} &\quad \xi_T\ll d \\
\label{eq:cleanvalid2}
l\gg d \quad & \text{if} & \quad \xi_T\gg d\; .
\end{eqnarray}
We note that this includes the region $d\ll l \ll \xi_T$, the finite
thickness preventing the small mean free path $l \ll \xi_T$ of
becoming effective. In the remaining part of the ballistic regime, 
see Fig.~\ref{fig:regime}, the full solution is not known,
but we may produce an approximate solution, which characterizes well
the numerical results found below. Limiting ourselves to $\xi_T \ll d$
allows us to consider the Green's function for the first Matsubara
frequency $\omega=\pi T$ only. We restrict ourselves to the forward
direction $v_{\text{x}}=+v_{\text{F}}$. From Eq.~(\ref{eilenreal}) we
find that $f_0\approx 2 \exp(-x/\xi_T)$ remains unchanged as in
(\ref{cleanreal}), and $f_0^{\dag}\ll 1$ and $1-g_0\ll 1$ obey the
approximate equations,
\begin{eqnarray}
 \label{eq:cleancorrection}
 \left(\frac{d}{dx}-\frac{1}{\xi_T}\right)f_0^\dagger(x)&=&-\frac1l\langle
 f_0(x)\rangle \\\nonumber \frac{d}{dx}(1-g_0(x))&=&-\frac1l\langle f_0(x)
 \rangle f(x)\;
\end{eqnarray}
with the approximate solutions
\begin{eqnarray}
 \label{eq:clean_corr_solu_for_f}
 f_0^\dagger(x)&=&\frac{\xi_T}{2l}e^{\displaystyle-x/\xi_T}\\
 \label{eq:clean_corr_solu_for_g}
 1-g_0(x)&=&\frac{\xi_T}{2l}e^{\displaystyle-2x/\xi_T}\; 
\end{eqnarray}
Here we have used $\langle f_0\rangle\approx f/2$, which is valid since
$f_0(-|v_{\text{x}}|)\ll f_0(|v_{\text{x}}|)$. Interestingly, while the
induced superconducting correlations as described by $\langle
f_0(x)\rangle$ remain unchanged as compared to the clean limit
(\ref{cleanreal}), the values of $f_0^{\dag}$ and $1-g_0$ are of order
$\xi_T/l$ rather than exponentially suppressed as in
(\ref{cleanreal}).  This is of importance for the current response as
we show below.

\subsection{Dirty limit}
If impurity scattering dominates, as described by $\langle
g_0\rangle/\tau\gg\omega$ and $\langle f_0\rangle /\tau\gg \Delta$,
Eq.~(\ref{eilenreal}) can be reduced to the Usadel equation
\cite{usadel} for the isotropic part $\langle f_0(x)\rangle$.  Assuming
$\omega\ll\Delta$ the solution in the normal metal takes the form
\begin{equation}
 \label{eq:dirtysolution}
 \langle f_0(x) \rangle =
\cosh(\sqrt{\frac{2\omega}{D}}(d-x))/\cosh(\sqrt{\frac{2\omega}{D}}d)\;,
\end{equation}
where $D=v_{\text{F}}^2\tau/3$ is the diffusion constant.  Equation
(\ref{eq:dirtysolution}) shows that the important energy scale is the
Thouless energy $E_{\text{Th}}=D/2\pi d^2$. The coherence length in
this case is $\xi_D(T)=(D/2\pi T)^{1/2}$, which reflects the diffusive
nature of the electron motion.

In the normal metal $l\ll\xi_T,d$ are necessary conditions for the
Usadel theory to be valid. However, as the numerical results will
confirm below, the Usadel theory in the normal metal may not be
applied without considering the superconductor inducing the proximity
effect.  The application of the Usadel equations requires the Green's
functions to be nearly isotropic, which in the superconductor is only
fulfilled for $\langle f_0\rangle/\tau\gg \Delta$. The validity of the
Usadel theory (in the absence of fields) is thus restricted to $l\ll
d,\xi_T$, and $l\ll \xi_0=v_{\text{F}}/2\pi T_c$, the dirty limit, see
Fig.~\ref{fig:regime}.

\subsection{Intermediate diffusive regime}

\begin{figure}[t]
 \begin{center}
 \leavevmode
 \psfig{figure=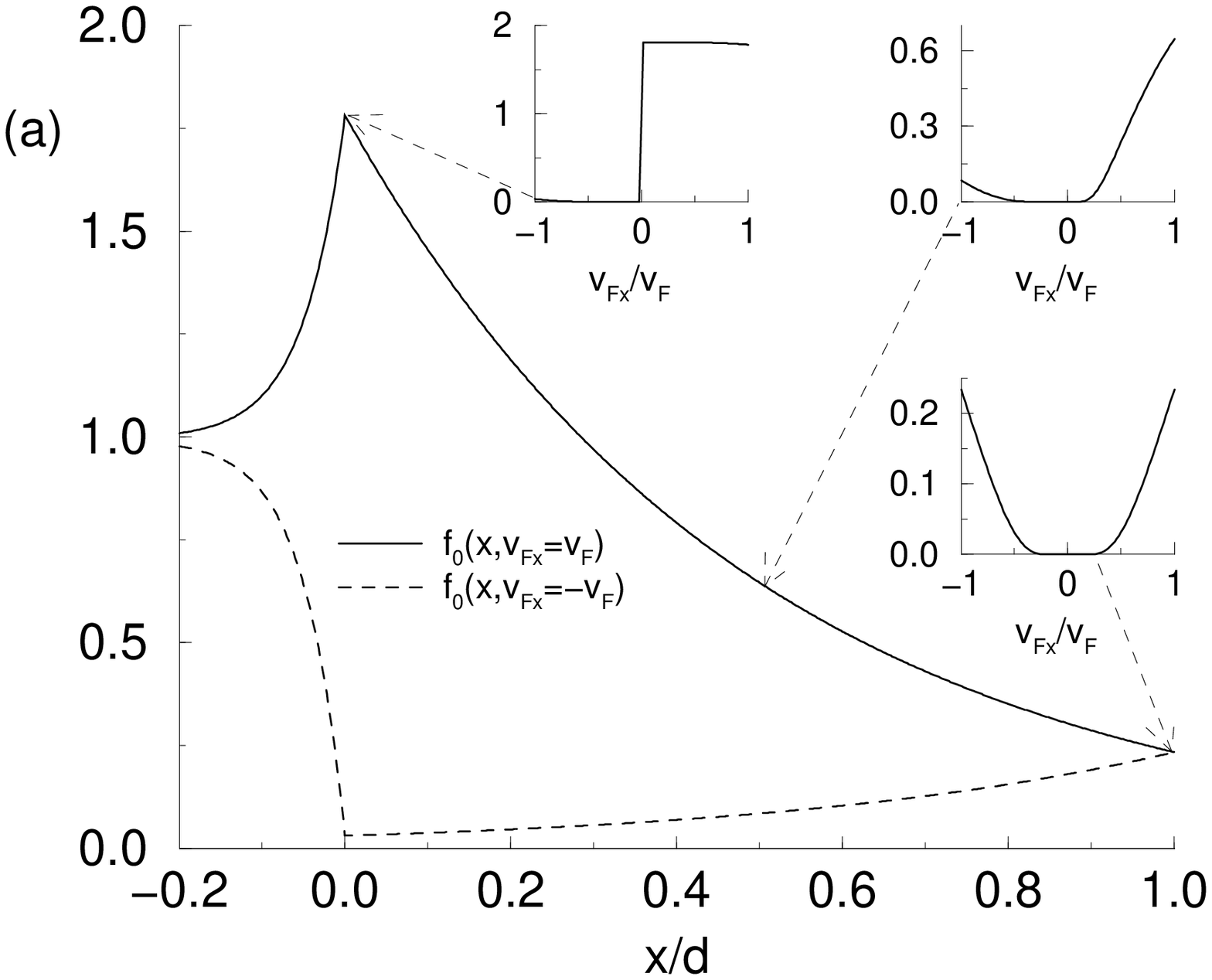,width=7cm}
 \psfig{figure=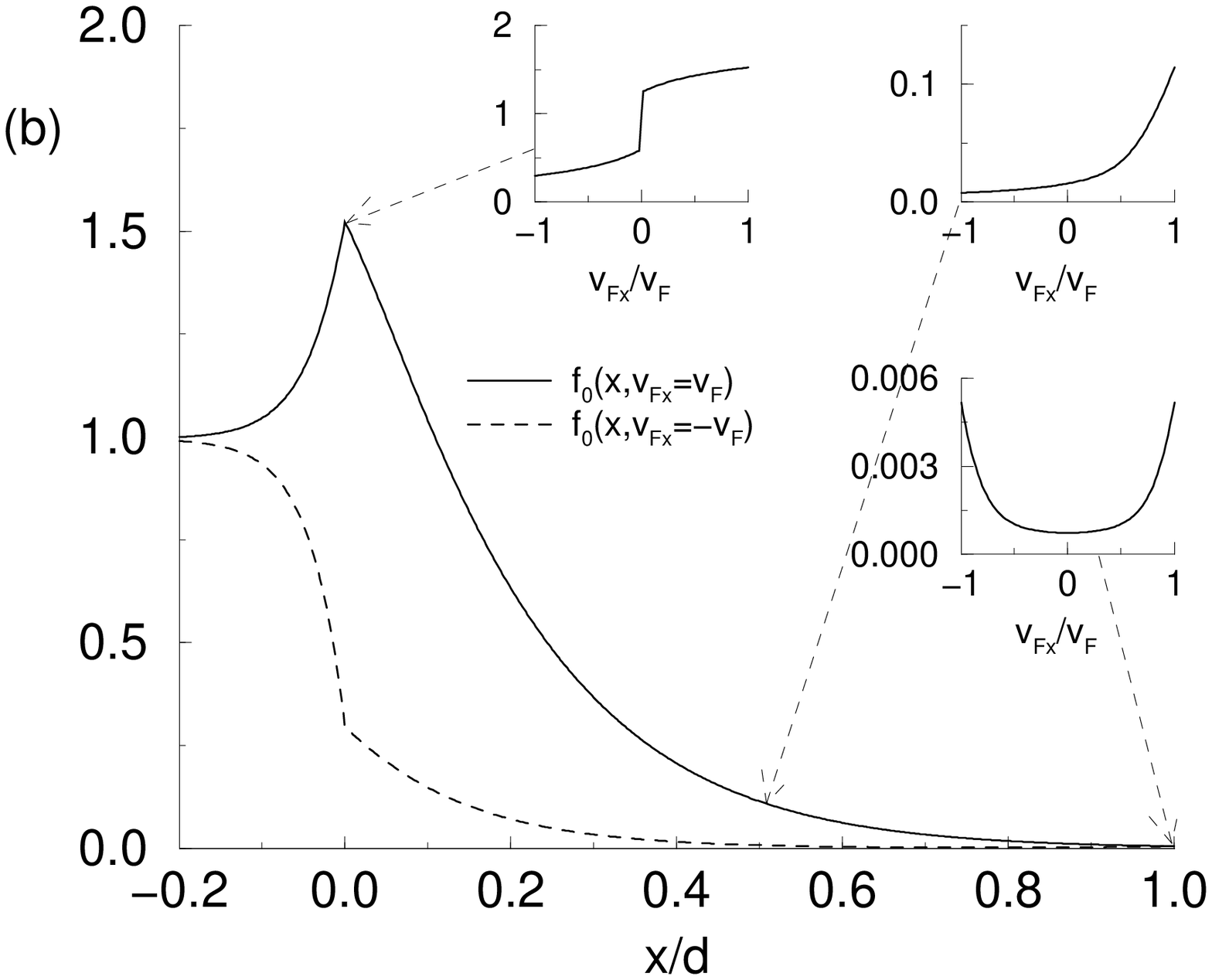,width=7cm}
 \caption[]{
   Spatial dependence of the anomalous Green's function $f_0(x,v_x)$
   in a proximity normal-metal layer.  The thickness is $d=10
   v_{\text{F}}/T_c$ and the frequency is $\omega=v_{F}/d$. The mean
   free path is $l=d$ in (a) and $l=0.1 d$ in (b). The insets show the
   angular dependence $f(v_{\text{Fx}})$ at the positions indicated by
   the arrows. We note that in the Usadel theory the angular
   dependence would be given by a linear function.}
 \label{fig:fvxtau1}
 \end{center}
\end{figure}

Now we relax all restrictions on the mean free path and investigate
the regime between the ballistic regime and the dirty limit.
Eq.~(\ref{eilenreal}) can be formally decoupled using the
Schopohl-Maki transformation \cite{schopohl}
\begin{equation}
 \label{schopohltrafo}
 a_{0}(v_{\text{x}},x) =
 \frac{f_{0}(v_{\text{x}},x)}{1+g_{0}(v_{\text{x}},x)},
 a^{\dagger}_{0}(v_{\text{x}},x) =
 \frac{f^{\dagger}_{0}(v_{\text{x}},x)}{1+g_{0}(v_{\text{x}},x)},
\end{equation}
leading to the Riccati differential equations
\begin{eqnarray}
  \nonumber
 %\lefteqn{
   -{v}_{x}\frac{d}{dx}a_{0}(v_{\text{x}},x)=%}
 % \\\nonumber & & 
 2\tilde{\omega}(x)a_{0}
 (v_{\text{x}},x)+\tilde{\Delta}(x)(a^2_0(v_{\text{x}},x)-1)
 \\\label{riccatireal}
 %\lefteqn{
   v_{\text{x}}\frac{d}{dx}a^\dagger_{0}(v_{\text{x}},x)=%}
 %\\\nonumber & & 
 2\tilde{\omega}(x)a^\dagger_{0}(v_{\text{x}},x)+\tilde{\Delta}(x)
 (a^{\dagger 2}_0(v_{\text{x}},x)-1).
\end{eqnarray}
Eqs.~(\ref{riccatireal}) provide the basis for a (stable) numerical
solution. We have determined the impurity self-energies
self-consistently by an iteration procedure starting from the
dirty-limit expression. Representative results of the numerical
calculation are shown in Fig.~\ref{fig:fvxtau1}. 

We have chosen the frequency $\omega=v_{F}/d$ and a mean free path of
$l=d$ in Fig.~\ref{fig:fvxtau1}(a), $l=0.1 d$ in
Fig.~\ref{fig:fvxtau1}(b). Note the distinction of the $f$-function
for forward ($v_{\text{Fx}}=v_{\text{F}}$, solid line) and backward
propagation ($v_{\text{Fx}}=-v_{\text{F}}$, dashed line). As we cross
over from the ballistic to the diffusive regime, the backward
propagating branch changes from a monotonically increasing
$f$-function of $x$ for $l=d$ to a decaying $f$-function for $l=0.1
d$. In the ballistic case the backward moving electrons carry
superconducting correlations only after reflection from the
normal-metal--vacuum boundary $x=d$. In the diffusive case, the
backward propagating $f$-function is generated by the impurity
scattering from the forward branch, thus taking the same functional
dependence of $x$, see Fig.~\ref{fig:fvxtau1}. This behavior is
illustrated in the insets, where $f$ is plotted as a function of
$v_{\text{x}}$ for fixed positions. The sharp features present in
Fig.~\ref{fig:fvxtau1}(a) are washed out by impurity scattering in
\ref{fig:fvxtau1}(b). Remarkably, however, they are still far from the
dirty-limit behavior, for which $f_0(v_{\text{Fx}})$ in the insets is
expected to be a straight line. According to conventional wisdom
$\l=0.1 v_{\text{F}}/\omega \ll \xi_T$ would indicate the dirty limit.
Considering that we have chosen $d=10v_{\text{F}}/T_c$, implying
$l\approx v_{\text{F}}/\Delta$, we notice that the dirty-limit
condition is not fulfilled in the superconductor. The anisotropy of
the $f$-function present in the superconductor by proximity induces an
anisotropy inside the normal metal, which is not accurately described
by the Usadel theory. For the dirty-limit theory to be valid in the
normal metal, the superconductor has to be dirty as well.

We note that in our present calculation we have assumed the same mean
free path in the superconductor and the normal metal. Allowing for
different mean free paths would affect the definition of the dirty and
the intermediate regime, and we do not further investigate this
question here.

\section{Linear-response kernel}
\label{sec:linmagresponse}
In this section we derive the general linear-response kernel
(\ref{kernel}) of a normal-metal--superconductor sandwich in terms of
the Green's functions in absence of the fields. We consider the
quasi-one-dimensional system shown in Fig.~\ref{fig:geometry},
assuming a superconductor of thickness $d_s$ and a normal metal of
thickness $d$. The magnetic field is applied in $z$-direction as
described by the gauge $\bbox{A}=A(x)\bbox{e}_{y}$. To calculate the
linear diamagnetic response, we separate the Green's functions into
its real and imaginary parts, where the imaginary part is of first
order in $A$ and the real part of zeroth order:
\begin{eqnarray}
 \label{expand}
 f(\bbox{v}_{\text{F}},\bbox{x})& = & 
 f_{0}(v_{\text{x}},x)+i f_{1}(v_{\text{x}},v_{\text{y}},x)
 \nonumber\\
 f^{\dagger}(\bbox{v}_{\text{F}},\bbox{x}) &=&
 f^{\dagger}_{0}(v_{\text{x}},x)+ i 
 f^{\dagger}_{1}(v_{\text{x}},v_{\text{y}},x) 
 \\\nonumber
 g(\bbox{v}_{\text{F}},\bbox{x}) &=& g_{0}(v_{\text{x}},x)+ i
 g_{1}(v_{\text{x}},v_{\text{y}},x)\; .
 \nonumber
\end{eqnarray}
The zeroth-order parts obey Eq.~(\ref{eilenreal}) discussed in
the previous section. The first-order parts of
Eq.~(\ref{eilenberger}) read
\begin{eqnarray}
 \label{eilenfirst}
 \lefteqn{-{v}_{x}\frac{d}{dx}f_{1}(v_{\text{x}},v_{\text{y}},x)=
 2\tilde{\omega}(x)f_{1}(v_{\text{x}},v_{\text{y}},x)}
 \\\nonumber 
 & & -2\tilde{\Delta}(x)g_{1}(v_{\text{x}},v_{\text{y}},x)
 +2ev_{\text{y}}A(x) f_{0}(v_{\text{x}},x) 
 \\\nonumber
 \lefteqn{{v}_{x}\frac{d}{dx}f^{\dagger}_{1}(v_{\text{x}},v_{\text{y}},x)=
 2\tilde{\omega}(x)f^{\dagger}_{1}(v_{\text{x}},v_{\text{y}},x)}
 \\\nonumber & & -2\tilde{\Delta}(x)g_{1}(v_{\text{x}},v_{\text{y}},x)
 +2ev_{\text{y}}A(x) f^{\dagger}_{0}(v_{\text{x}},x)\; .
\end{eqnarray}
where $\tilde{\omega}(x)$, $\tilde{\Delta}(x)$ were given after 
Eq.~(\ref{eilenreal}) and 
\begin{eqnarray}
 \label{g1}
 \lefteqn{g_{1}(v_{\text{x}},v_{\text{y}},x)=}\\\nonumber
 && -\frac{f_{0}(v_{\text{x}},x)f^{\dagger}_{1}(v_{\text{x}},v_{\text{y}},x)
 +f_{1}(v_{\text{x}},v_{\text{y}},x)f^{\dagger}_{0}(v_{\text{x}},x)}{
 2g_{0}(v_{\text{x}},x)}
\end{eqnarray}
follows from the normalization (\ref{norm}). We now apply the
Maki-Schopohl transformation defined in (\ref{schopohltrafo}) to the
full equations of motions (\ref{eilenberger}). After linearization we
obtain
\begin{eqnarray*}
 f_{1}(v_{\text{x}},v_{\text{y}},x)&=&
 2\frac{a_{1}(v_{\text{x}},v_{\text{y}},x)-
 a_{0}^{2}(v_{\text{x}},x)
 a^{\dagger}_{1}(v_{\text{x}},v_{\text{y}},x)}{
 (1+a_{0}(v_{\text{x}},x)a^{\dagger}_{0}(v_{\text{x}},x))^{2}}\\
 f^{\dagger}_{1}(v_{\text{x}},v_{\text{y}},x)&=&
 2\frac{a^{\dagger}_{1}(v_{\text{x}},v_{\text{y}},x)
 -a^{\dagger2}_{0}(v_{\text{x}},x)
 a_{1}(v_{\text{x}},v_{\text{y}},x)}{
 (1+a_{0}(v_{\text{x}},x)a^{\dagger}_{0}(v_{\text{x}},x))^{2}}\; ,
\end{eqnarray*}
and Eqs.\ (\ref{eilenfirst}) are decoupled into
\begin{eqnarray}
 \lefteqn{-\frac{v_{\text{x}}}{2}\frac{d}{dx}
   a_{1}(v_{\text{x}},v_{\text{y}},x)=}
 \label{riccatifirst}
 & &\\\nonumber \quad & &
 \left[ \tilde{\omega}(x)+
 \tilde{\Delta}(x)a_{0}(v_{\text{x}},x)\right]
 a_{1}(v_{\text{x}},v_{\text{y}},x)+ 
 ev_{\text{y}}A(x) a_{0}(v_{\text{x}},x)\\\nonumber
 \lefteqn{\frac{v_{\text{x}}}{2}\frac{d}{dx}a^{\dagger}_{1}
 (v_{\text{x}},v_{\text{y}},x)=} 
 \\\nonumber \quad & & 
 \left[ \tilde{\omega}(x)+
 \tilde{\Delta}(x)a^{\dagger}_{0}(v_{\text{x}},x)\right]
 a^{\dagger}_{1}(v_{\text{x}},v_{\text{y}},x)+
 ev_{\text{y}}A(x) a^{\dagger}_{0}(v_{\text{x}},x).
\end{eqnarray}
For a more general form of these equations that has been used to treat
the linear electromagnetic response of vortices numerically, see
\onlinecite{eschrig}. We will now proceed analytically.  As a
consequence of Eq.~(\ref{riccatifirst}) we find
$a_{1}(-v_{\text{y}})=-a_{1}(v_{\text{y}})$ and the same for
$a^{\dagger}_{1}$, which leads to $\langle f_{1}\rangle =\langle
f^{\dagger}_{1}\rangle =0$, as was already noted above. Furthermore,
since $a^{\dagger}_{1}(v_{\text{x}})=a_{1}(-v_{\text{x}})$, we only
have to consider one of the two equations (e.g., the first one).
Equation (\ref{riccatifirst}) is an inhomogeneous first-order
differential equation, which can be integrated analytically. Assuming
that $f$ and $f^{\dag}$ do not change sign with the help of
(\ref{eilenreal}) the solution can be written as
\begin{eqnarray}
 \label{solution}
 \lefteqn{a_{1}(v_{\text{x}},v_{\text{y}},x)= c(v_{\text{x}},v_{\text{y}})
 \frac{m(v_{\text{x}},x,x_{0})}{f^{\dagger}_{0}(v_{\text{x}},x)}}
 \\ \nonumber& & 
 -\frac{2ev_{\text{y}}}{v_{\text{x}}f^{\dagger}_{0}(v_{\text{x}},x)}
 \int_{x_{0}}^{x} [1-g_{0}(v_{\text{x}},x^{\prime})]
 m(v_{\text{x}},x,x^{\prime}) 
 A(x^{\prime})dx^{\prime}
\end{eqnarray}
where 
\begin{equation}
 \label{defm}
 m(v_{\text{x}},x,x^{\prime}) = 
 \exp\left(\frac{2}{v_{\text{x}}}\int_{x}^{x^{\prime}}
 \frac{\tilde{\Delta}(x^{\prime\prime})}{
 f^{\dagger}_{0}(v_{\text{x}},x^{\prime\prime})}dx^{\prime\prime}\right)
\end{equation}
In this equation $x_{0}$ is an arbitrary reference point and the
constant $c$ has to be determined by the appropriate boundary
conditions. $m$ satisfies the relations of a propagator, $
m(u,x,x^{\prime}) = m(u,x^{\prime},x)^{-1}$ and $
m(u,x,x^{\prime\prime}) m(u,x^{\prime\prime},x^{\prime})=
m(u,x,x^{\prime})$.  Now we determine the constant $c$ for a system of
size $[-d_s,d]$. We assume specular reflection at two boundaries at
$x=-d_s,d$ and ideal interfaces between different materials inside the
system. The appropriate boundary conditions are
$f(v_{\text{x}},v_{\text{y}},x=-d_{s},d) =
f(-v_{\text{x}},v_{\text{y}},x=-d_{s},d)$ and continuity at the
internal interfaces. The same conditions are valid for $a_{1}$ and
$a^{\dagger}_{1}$. This leads to
\begin{eqnarray}
 \label{constant}
 c(v_{\text{x}},v_{\text{y}}) &= &
 2e\frac{v_{\text{y}}}{v_{\text{x}}}\int_{-d_{s}}^{d}
 \frac{m(v_{\text{x}},d,x^{\prime})+m(-v_{\text{x}},d,x^{\prime})}{ 
 m(v_{\text{x}},d,-d_{s})-m(-v_{\text{x}},d,-d_{s})}\nonumber\\&&
 \times [1-g_{0}(v_{\text{x}},x^{\prime})] 
 A(x^{\prime})dx^{\prime}\;\hfill .
\end{eqnarray}
The current is determined by Eq.~(\ref{current}), using the Green's
functions (\ref{g1}), expressed by the solution (\ref{solution}). We
obtain the following general result for the linear current response in
functional dependence of the vector potential,
\begin{equation}
 \label{jy}
 j_{y}(x) = -\int\limits_{-d_{s}}^{d} K(x,x^\prime)
 A(x^{\prime})dx^{\prime}\; ,
\end{equation}
where the kernel $K(x,x^{\prime})$ is given by
\end{multicols}
\begin{eqnarray}
 \label{kernel}
 K(x,x^\prime) &=& 
 \frac{e^{2}p_{F}^{2}}{\pi}T\sum_{\omega>0}
 \int\limits_{0}^{v_{F}}\text{d}u\, \frac{v_{F}^{2}-u^{2}}{{v_{F}^{2}}u}
 [1+g_{0}(u,x)][1-g_{0}(u,x^{\prime})]
 \bigg[\Theta(x-x^{\prime})m(u,x,x^{\prime})+
 \Theta(x^{\prime}-x)m(-u,x,x^{\prime})\\
 && \nonumber
 +\frac{ m(-u,x,d)m(u,d,x^{\prime})}{ 1- m(u,d,-d_s) m(-u,-d_s,d) }
 + \frac{m(u,x,-d_s)m(-u,-d_s,x^{\prime})}{ 1- m(u,d,-d_s) m(-u,-d_s,d) }\\
 && \nonumber
 +\frac{m(-u,x,d)m(u,d,-d_s)m(-u,-d_s,x^{\prime})}{1-m(u,d,-d_s) m(-u,-d_s,d)}
 +\frac{m(u,x,-d_s)m(-u,-d_s,d)m(u,d,x^{\prime})}{1-m(u,d,-d_s) m(-u,-d_s,d) } 
\bigg]\; .
\end{eqnarray}
\widetext
\begin{multicols}{2}

Equation (\ref{kernel}) gives the exact linear-response kernel of
any quasi-one-dimensional system, consisting of a combination of
normal and superconducting layers extending from $x=-d_s$ to $x=d$.
The kernel is expressed in terms of the quasi-classical Green's
functions in absence of the fields, which may be specified for the
particular problem of interest. We note two characteristic features
of Eq.~(\ref{kernel}): The factor $(1-g_{0}(u,x^{\prime}))$ measures
the deviation of a quasi-classical trajectory from the normal state
$g_0\equiv 1$, which is inert to a magnetic field. The propagator
$m(u,x,x^{\prime})$ shows up in six summands which represent all the
ballistic paths from $x$ to $x^{\prime}$, accounting for multiple
reflection at the walls at $-d_s$ and $d$. Thus the first two
summands connecting $x$ and $x^{\prime}$ directly constitute the
bulk contribution, while the additional four summands are specific
to a finite system (assuming specular reflection at the boundary).
We note that a form similar to (\ref{kernel}) may be derived for
non-ideal interfaces between the normal and superconducting layers,
if the appropriate boundary conditions following from Ref.\ 
\onlinecite{zaitsev} are taken into account (these boundary
conditions are only valid if the distance between two barriers is
larger than the mean free path).

For illustration we reproduce the current response of a half-infinite
superconductor. Setting $d=0$ and $d_s\to\infty$, the solution of the
Eilenberger equation (\ref{eilenreal}) takes the simple form
$g_0=\omega/\Omega$, $f_0=f^{\dag}_0=\Delta/\Omega$, where
$\Omega=(\Delta^2+\omega^2)^{1/2}$. Inserting in (\ref{kernel}) we
obtain the linear-response kernel
\begin{eqnarray}
 \label{kernelsuper} 
 K_{S}(x,x^{\prime}) & =&\frac{e^{2}p_{F}^{2}}{\pi}
 T\sum_{\omega>0}\frac{\Delta^{\!2}}{\Omega^2}
 \int\limits_{0}^{v_{F}}\!\!du\frac{1\!-\!u^{2}/v_{F}^{2}}{u}
 \\\nonumber&&\times\big[
 e^{-(2\Omega+\frac{\scriptstyle 1}{\scriptstyle\tau}) 
 \frac{\scriptstyle|x-x^{\prime}|}{\scriptstyle u}}\!\!+
 e^{(2\Omega+\frac{\scriptstyle 1}{\scriptstyle\tau})
 \frac{\scriptstyle x+x^{\prime}}{\scriptstyle u}}\big]\; ,
\end{eqnarray}
which describes the current response of an arbitrary superconductor,
as first derived by Gorkov\cite{gorkov}, which here additionally
includes the effect of the boundary. For fields varying rapidly
spatially we arrive at a non-local current-field relation of the
Pippard-type\cite{pippard}, while for slowly varying fields the kernel
can be integrated out in Eq.~(\ref{jy}), producing the London
result\cite{london}. We recall here certain generic features of this
kernel, which are of importance below. In a dirty superconductor
($\Omega\ll 1/\tau$) the range is given by the mean free path
$l=v_{F}\tau$. In a clean superconductor ($1/\tau\ll\Omega$), the
range is roughly given by the coherence length $\xi_0$ and is thus
nearly temperature-independent.

\section{Magnetic response}
\label{sec:magresponse}

For the NS system we consider in this paper, see
Fig.~\ref{fig:geometry}, the kernel (\ref{kernel}) may be simplified
using $m(u,x,-\infty) \rightarrow 0$ and $m(-u,-\infty,x) \rightarrow
0$ as $d_s\to \infty$ ($u>0$). The linear-response kernel takes the
form,
\end{multicols}
\begin{eqnarray}
 \label{jyinf}
 K(x,x^\prime)& =& -\frac{e^{2}p_{F}^{2}}{\pi}T\sum_{\omega>0}
 \int\limits_{0}^{v_{F}}du\frac{v_{F}^{2}-u^{2}}{v_{F}^{2}u}
 (1+g_{0}(u,x)) (1-g_{0}(u,x^{\prime}))\\
 \nonumber & &
 \bigg[\Theta(x-x^{\prime})m(u,x,x^{\prime})+
 \Theta(x^{\prime}-x)m(-u,x,x^{\prime})
 + m(-u,x,d)m(u,d,x^{\prime}) \bigg]\;.
\end{eqnarray}
\widetext
\begin{multicols}{2}
The magnetic response of the proximity system follows from the 
self-consistent solution of Eq.~(\ref{jy}) and the Maxwell equation
\begin{equation}
 \label{maxwell}
 \frac{d^{2}}{dx^{2}}A(x)=-4\pi j_{y}(x)\; .
\end{equation}
As boundary condition we use $\frac{d}{dx}A(x)|_{x=d}=H$, where $H$ is
the applied magnetic field, and $A(0)=0$, neglecting the penetration
of the field into the superconductor. The inclusion of the field
penetration into the superconductor leads to corrections $\sim
\lambda_{\text{S}}/d$ to $\rho$, which is a small ratio for typical
proximity systems. Here $\lambda_{\text{S}}$ is the effective
penetration depth of the superconductor, including the nonlocal or
impurity effects. The magnetic response of the normal-metal layer is
measured by the screening fraction $\rho=-4\pi\chi=1- A(d)/Hd$, which
gives the fraction of the normal-metal layer that is effectively field
free. It is given by the susceptibility $\chi$, which is equal to the
ratio of the average magnetization to the applied magnetic field.

The general properties of the kernel (\ref{kernel}) are characterized
by both the decay (range) of the propagator $m(v_{F},x,x^{\prime})$
and the amplitude of the prefactor $(1+g_0)(1-g_0)$ which determines
the degree of non-locality of the relations (\ref{jy}). The inverse
decay length of the propagator is proportional to the off-diagonal
part of the self-energy $\tilde{\Delta}$ and the prefactor is related
to the superfluid density. We discuss below how, in the proximity
effect, the range of the kernel varies from infinity to $l$ and
$\xi_T$, exhibiting a strong temperature dependence, which leads to
non-trivial screening properties. Furthermore, the superfluid density
introduces an additional length scale in the problem: the London
length $\lambda_N$, which becomes crucial for the distinction of
various regimes.

\subsection{Clean limit}

A special case is the clean normal metal ($l\to \infty$). Here the
range is infinite and the current-field relation is completely
non-local. It follows from (\ref{kernel}) that it is necessary to have
impurities in a normal metal to get a finite range of the kernel. In
the limit $d\gg\xi_0$ the current may be written as
\begin{equation}
 \label{eq:jclean}
 j_{\text{clean}}=-\frac{1}{4\pi\lambda^{2}(T)\,d}
 \int\limits_{0}^{d}A(x)dx \; .
\end{equation}
This defines a temperature-dependent penetration depth that can be given
explicitly in the limits $T=0$ and $T\gg T_A$:
\begin{equation}
 \label{lambdaclean}
 \lambda^{2}(T)=
 \left\{
 \begin{array}[c]{ll}\displaystyle
 \frac{m}{4\pi e^{2}n_{e}}=:\lambda_N^{2} &;\; T=0
 \\[5mm]\displaystyle
 \frac{\lambda_N^{2} T}{12T_A}e^{2\frac{T}{T_A}}&;\; T\gg T_{\text{A}}.
 \end{array}\right.
\end{equation}
Solving Maxwell's equation we find
\begin{equation}
 \label{aclean}
 A(x)=Hx(1-\frac{3}{4}\frac{d(2d-x)}{(3 \lambda^2(T)+d^2)}),
\end{equation}
and the screening fraction
\begin{equation}
 \label{rhoclean}
 \rho=\frac{3}{4+12\lambda^{2}(T)/d^{2}}\; .
\end{equation}
In the limit $\lambda(T)\ll d$ the screening fraction is $3/4$, thus
the screening is not perfectly diamagnetic. The magnetic field inside
the normal metal is $B(x)/H=1-2\rho(1-x/d)$, showing the effect of
overscreening for $\rho>1/2$, where the field reverses sign inside the
normal metal.

\subsection{Dirty limit}

Using the fact that the zeroth-order Green's function is nearly
isotropic and varies on a scale $\xi_D(T)\gg l$, we find for the
kernel (\ref{jyinf})
\begin{eqnarray}
 \label{eq:kerneldirty}
 K(x,x^\prime)&=&-\frac{e^{2}p_{F}^{2}}{\pi}T\sum_{\omega>0}
 \langle f(x)\rangle^2
 \int\limits_{0}^{v_{F}}du\frac{v_{F}^{2}-u^{2}}{v_{F}^{2}u}
 \\\nonumber &&\times
 \left[e^{-\frac{|x-x^\prime|}{l u}} + 
 e^{\frac{ 2d-x-x^\prime}{l u}}\right]\; .
\end{eqnarray}
The kernel is factorized in a part containing the temperature
dependence and a part which is responsible for the non-locality. The
current is then expressed as
\begin{equation}
 \label{eq:currentdirty}
 j(x)=-\frac{1}{4\pi \lambda^2(x,T)} 
 \int\limits_0^d K_d(x,x^\prime)A(x^\prime)dx^\prime\; .
\end{equation}
The local penetration depth $\lambda(x,T)$ is defined as
\begin{equation}
 \label{eq:localpendepth}
 \frac{1}{\lambda^2(x,T)}=\frac{4\pi\tau}{\lambda_N^2} 
 T\sum_{\omega>0} \langle f(x)\rangle^2
\end{equation}
and the temperature-independent part of the kernel is given by
\begin{eqnarray}
 \label{eq:kerneldirty2}
 K_d(x,x^\prime)&=&\frac{3}{4l}\left[
 E_1(\frac{|x-x^\prime|}{l})-E_3(\frac{|x-x^\prime|}{l})\right.
 \\\nonumber
 & & \left.+E_1(\frac{2d-x-x^\prime}{l})-E_3(\frac{2d-x-x^\prime}{l})
 \right]\; .
\end{eqnarray}
In this formula $E_{n}(z)=\int_1^{\infty} t^{-n}\exp(-zt)dt$ is the
exponential integral. For $\lambda(x,T)\gg l$ the vector potential may
be taken out of the integral in Eq.~(\ref{eq:currentdirty}) and the
spatial integral yields the well-known local current-vector potential
relation used in Usadel theory\cite{usadel}. We note that for
$\lambda\ll\xi_0$ there may exist a region (see Fig.~\ref{fig:regime})
where the Green's functions are nearly isotropic, and in absence of
the field are given by Usadel theory, but the current response is
nonlocal.  To put limits on the validity of the local relation, we
consider the approximate form $\langle
f\rangle\sim\exp[-x(2\omega/D)^{1/2}]$ to determine the local
penetration depth. As a result we find
\begin{equation}
 \label{eq:usadelpendepth}
 \lambda(x,T)\approx
 \left\{
 \begin{array}[c]{ll}
 \lambda_N\frac{x}{l}& 
 \text{if}\quad\xi_D(T)\gg d\\
 \lambda_N\frac{\xi_D(T)}{l}e^{x/\xi_D(T)}&
 \text{if}\quad\xi_D(T)\ll d\;.
 \end{array}
 \right.
\end{equation}
To achieve locality we need to have $l<\lambda(x,T)$ in the region,
where the screening takes place. For $T\ll D/d^2$ this means
$l\ll\lambda(d)$ leading to the condition $l^2\ll \lambda_N d$. For
$T\gg D/d^2$ screening takes place at $x\approx\xi_D$ and we have
$l^2\ll\lambda_N\xi_D(T)$. The local penetration depth at the outer
boundary can be small compared to $d$. In that case the screening
fraction $\rho = 1-\lambda(d)/d$ can reach practically unity.

At high temperatures $T\gg E_{\text{Th}}$, the inverse penetration
depth is exponentially suppressed on a scale of the
temperature-dependent coherence length $\xi_D(T)$. This length defines
the screening region and consequently
\begin{equation}
 \label{eq:rhodirty1}
 \rho(T)\propto \frac{\xi_{D}(T)}{d} \propto T^{-1/2}\; .
\end{equation}
This result has already been obtained on the basis of Ginzburg-Landau
theory \cite{degennes} and numerically confirmed using Usadel theory
\cite{narikiyo:89}. We expect that non-local screening, which may be
taken into account using Eq.~(\ref{eq:kerneldirty2}), will only lead
to quantitative corrections to Eq.~(\ref{eq:rhodirty1}). We will not
consider this here but concentrate on the more interesting case in
which non-locality gives rise to a qualitatively different picture.

\subsection{Arbitrary impurity concentration: numerical results}

As has been shown in the last two subsections, there are two main
differences in the observable properties of the induced screening in
the clean or the dirty limit. Firstly, the saturation value of $\rho$
in the dirty limit can reach practically unity, whereas in the clean
limit it is limited to $3/4$. The analytic behavior at high
temperatures is quite different too. In the dirty limit $\rho$ shows
an algebraic behavior $\propto T^{-1/2}$, whereas in the clean limit
we find $\rho \propto \exp(-2 T/T_A)$. From a theoretical point of
view these two limits are characterized by a completely nonlocal
constitutive relation in the clean limit and a local relation in the
dirty limit. In this section we will investigate the magnetic response
in the regime between these two extreme cases.

To calculate the diamagnetic response, we have evaluated the integral
kernel in Eq.~(\ref{jyinf}) numerically using the results from the
Sec.~\ref{sec:proxieffect} and solving Maxwell's equation by a
finite-difference technique. Therefore, the parameters entering the
calculation are $l/d$ and $\lambda_N/d$, assuming $l>\xi_0$, and
$\xi_T/d$ giving the temperature dependence.

\begin{figure}[h]
 \begin{center}
 \leavevmode 
 \psfig{figure=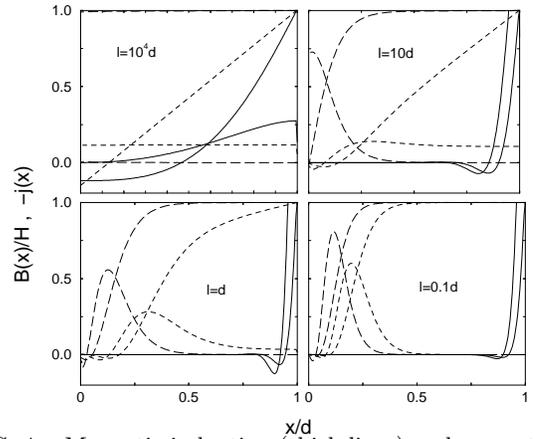,width=7cm}
\narrowtext
 \caption[]{
   Magnetic induction (thick lines) and current densities (thin lines) in
   a proximity layer for different mean free paths. The different
   curves in each graph correspond to temperatures of $T/T_A = 0.04$
   (solid line), $5$ (short-dashed line), and $8$ (long-dashed line).
   }
 \label{fig:field} 
 \end{center}
\end{figure}

Magnetic field distributions for various impurity concentrations and
temperatures are shown in Fig.~\ref{fig:field}. Thick lines show the
magnetic field and thin lines the current distribution inside the
normal metal. Different graphs correspond to different mean free paths
and the curves inside each graph to different temperatures. In all
curves we have chosen $\lambda_N=0.003 d$. All these curves clearly
show deviations from the clean-limit behavior, where the field decays
linearly and the current density is spatially constant. Larger
impurity concentrations make it possible to localize the current on a
length scale smaller than system size. Obviously, this scale is not
given by the mean free path, but can be considerably smaller. Later we
will show what determines this length scale. Whether the localization
of the current increases or decreases the screening fraction depends
on temperature. Note that for all parameters chosen the field is
overscreened, which is the signature of a nonlocal constitutive
relation.

\begin{figure}[h]
 \begin{center}
 \leavevmode 
 \psfig{figure=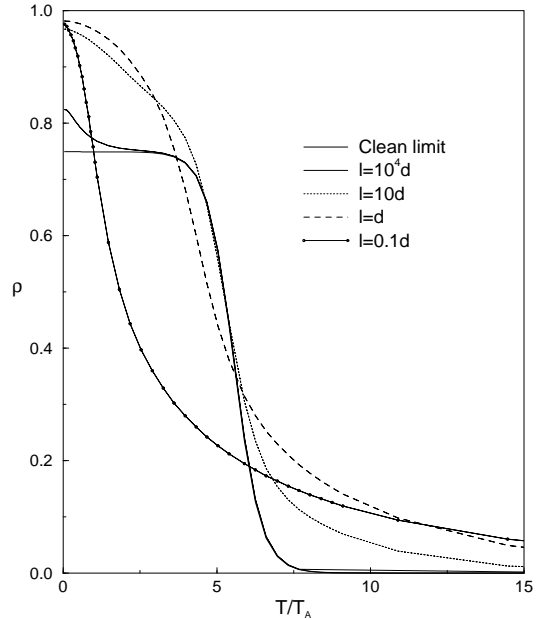,width=7cm}
 \caption[]{ 
   Numerical results for screening fraction of the normal metal layer
   for $\lambda_N=0.003d$. The clean limit is indicated by a thin line
   reaching $\rho=0.75$ for $T\to 0$. Even a very large mean free path
   of $10^4d$ leads to an enhanced screening at low temperatures. For
   smaller mean free paths (but still $>d$) the screening is enhanced
   at high and low temperatures. At the smallest mean free path $0.1
   d$ the system is in the diffusive regime leading to a completely
   different temperature dependence.}
 \label{fig:rho1} 
 \end{center}
\end{figure} 

The screening fraction as a function of temperature is shown in
Fig.~\ref{fig:rho1}. The different curves are for the clean limit and
for mean free paths $l/d=10^{4},10,1,0.1$. We see that a finite
impurity concentration has strong influence on the screening fraction,
even if $l>d$. It can either increase or decrease the diamagnetic
screening, depending on temperature.

For the interpretation of these results we first consider the case
$l=0.1 d$. The lower-right graph in Fig.~\ref{fig:field} shows that
the screening is nearly local, since overscreening is rather small.
The local screening strength depends on the local superfluid density.
At low temperatures $T\ll T_A$ the superfluid density at $x=d$ is
finite and the field is screened exponentially, leading to a screening
fraction of nearly unity. A higher temperature suppresses the
superfluid density and the field penetrates to the point where the
density is large enough to screen effectively. On the other hand, the
locality of the kernel allows the system to screen even if the
superfluid density is suppressed nearly everywhere. The screening is
then enhanced in comparison to the clean limit. In Fig.~\ref{fig:rho1}
this appears at $T\approx 6 T_A$.

Let us now consider a mean free path of order or much larger than the
sample size. Even for $l=10^4d$ we see a deviation from the
clean-limit expression at low temperatures. For $l=10d$ and $l=d$
screening is enhanced in comparison to the clean limit at low and high
temperatures. Only in an intermediate regime, i.~e. around $T=5T_A$ in
our case, $\rho$ is similar to the clean limit screening fraction.  A
qualitative understanding may be gained from looking at the
constitutive relation in the limit $l\gg d$. In the limit $T\ll T_A$
the zeroth-order Green's functions are given by the clean-limit
expressions (\ref{cleanreal}). We approximate the kernel (\ref{jyinf})
by
\begin{equation}
 \label{kernelapprox}
 K(x,x^\prime)=\frac{1}{8\pi \lambda^2(T)d} 
 \left[e^{-\frac{|x-x^\prime|}{l}}+
 e^{-\frac{2d-x-x^\prime}{l}}\right]\; .
\end{equation}
Since $l\gg d$, the exponentials may be expanded to first order. As a
result, we obtain two contributions to the current
\begin{eqnarray}
 \label{jyclean}
 j_{\text{clean}}&=&
 \frac{-1}{4\pi \lambda^2(T)d}\int_0^d A(x)dx\\
 &&\label{jyimp}\\\nonumber
 j_{\text{imp}}(x)&=&
 \frac{1}{8\pi \lambda^2(T)d}\int_0^d
 \frac{|x-x^\prime|+2d-x-x^\prime}{l}A(x^\prime)dx^\prime\; .
\end{eqnarray}
When will deviations from the clean limit become important? It is
clear that the impurities cannot be neglected, if (\ref{jyimp}) is
comparable to (\ref{jyclean}). We estimate this by calculating the two
contributions to the current using the clean-limit vector potential
(\ref{aclean}). Comparing the two contributions, we find that
impurities can be neglected, if
\begin{equation}
 \label{supercleancondition}
 \lambda_{\text{eff}}(T)\equiv\sqrt[3]{\lambda^2(T)l}\gg d\; .
\end{equation}

This equation defines a new length scale, the effective penetration
depth $\lambda_{\text{eff}}$, which determines the validity of the
clean-limit magnetic response. For the clean limit to be valid at
$T=0$ the condition $\lambda_{\text{eff}}(0)>d$ has to be fulfilled,
since in this case the screening takes place on the geometrical scale
$d$. In the case $\lambda_{\text{eff}}(0)\ll d$ the field is screened
on a scale $\lambda_{\text{eff}}$ and the screening fraction is
strongly enhanced in comparison to the clean limit. Nevertheless, the
clean-limit behavior reappears at higher temperatures, since
$\lambda_{\text{eff}}(T)$ grows with temperature.

For $T\gg T_A$ the deviations from the clean limit are related to
deviations of the zeroth-order Green's function from the clean-limit
expression due to impurity scattering. The correction to $g$, given in
Eq.~(\ref{eq:clean_corr_solu_for_g}), leads to a finite superfluid
density in the vicinity of the superconductor via the factor
$1-g(x^\prime)$ in the kernel. The range of the propagator is modified
by the correction (\ref{eq:clean_corr_solu_for_f}) to $f^\dagger$,
leading to
\begin{equation}
 \label{eq:mcorr}
 m(x,x^\prime)=\exp\left(\frac{1}{l}
 \int_x^{x^\prime}dx^{\prime\prime}\frac{\langle
 f(x^{\prime\prime})\rangle}{f^\dagger(x^{\prime\prime})}\right) 
\approx\exp\left(2\frac{x^\prime-x}{\xi_T}\right)\; .
\end{equation}
Thus, the range of the kernel is now given by $\xi_T$, which is
strongly temperature dependent. Summarizing, we find for the current
\begin{equation}
 \label{eq:currcorr}
 j(x)\approx \frac{-1}{\lambda_{\text{eff}}(0)^3}\int_0^d dx^\prime
 e^{-2\frac{x^\prime+|x-x^\prime|}{\xi_T}}A(x^\prime)\; ,
\end{equation}
again showing the importance of the new length scale
$\lambda_{\text{eff}}$. In the limit $\lambda_{\text{eff}}(0) \gg
\xi_T$ the field cannot be effectively screened on the scale $\xi_T$,
leading to a vanishing screening fraction. If $\lambda_{\text{eff}}(0)
\ll \xi_T$ the field can be screened on a length scale smaller than
$\xi_T$ and the screening fraction will be finite.
 
It is therefore evident that the interplay between
local and nonlocal physics is of crucial importance for the screening
behavior of a normal-metal proximity layer. The most interesting
regime occurs for $l>d$, where a transition between different
screening behaviors may be observed by varying the temperature, see
Fig.~\ref{fig:regime}.

We note that the screening fraction is a non-monotonic function of the
mean free path. At low temperature, with increasing mean free path
(i.e. increasing purity), the screening fraction is reduced rather
than enhanced. Assuming a temperature-dependent scattering mechanism
with decreasing mean free path as a function of temperature, such as
electron-electron or electron-phonon interaction, we might speculate
to observe a non-monotonic (i.e. re-entrant) behavior of the
susceptibility (here the smallness of the scattering rate is
compensated by the high sensibility of the non-local current-field
relation). However, as is evident from Eq.~(\ref{defm}), the largest
off-diagonal self-energy ($\tilde{\Delta}$) which includes e.g.
impurity scattering will provide a (low-temperature) cutoff for this
behavior.

Finally we comment on the effect of a rough boundary. For $T\gg T_A$
the Green's functions are independent of the boundary condition at
$x=d$. In this case a {\it finite} screening fraction can only be due
to impurity scattering, see Fig.~\ref{fig:rho1}. For $T<T_A$ the
screening behavior will be strongly affected by a rough boundary. This
makes it possible to distinguish between clean samples with a rough
boundary and samples containing impurities.

\section{Conclusions}
We have investigated the diamagnetic response of a proximity layer for
arbitrary impurity concentration using the quasiclassical theory of
superconductivity. We found a variety of different regimes in which
the physics is different from the previously studied clean and dirty 
limits.

We have first investigated the proximity effect in the absence of
fields, distinguishing three different regimes, see
Fig.~\ref{fig:regime}. In the ballistic regime, the
validity of the clean-limit solution is restricted to $l\gg
d\exp(2d/\xi_T)$ for $\xi_T\ll d$ and to $l\gg d$ for $\xi_T\gg d$.
The last condition is the consequence of the suppression of the
density of states for $\omega\ll v_{\text{F}}/d$, which enhances the
effective mean free path to $\sim l v_{\text{F}}/\omega d$. In the
diffusive regime, we found that the validity of the Usadel equation
(dirty limit) depends on the superconductor as well as the normal
metal, and is thus restricted to $l\ll \xi_0$ and $l\ll d,\xi_T$. The
first condition is due to the fact that 
the induced superconducting correlations are strongly anisotropic
for a clean superconductor ($l\gg \xi_0$), even if the motion is diffusive 
in the normal metal. The
intermediate diffusive regime ($\xi_0\ll l\ll d$) is not covered by
these two cases. Here the full Eilenberger equation has to be solved,
which requires a numerical analysis, see Fig.~\ref{fig:fvxtau1}.

To study the magnetic response of the proximity layer we have derived
explicit expressions for the general linear-response kernel (\ref{kernel}) 
for an NS sandwich.
%The current density $j(x)$ is given by a convolution of the kernel
%$K(x,x^\prime)$ with the vector potential $A(x^\prime)$, the kernel
%being determined by the Green's functions in the absence of fields.
This derivation may easily be generalized to systems such as Josephson
junctions or unconventional superconductors.
We have used this linear-response kernel to study the magnetic response
of the proximity system at arbitrary impurity concentrations. The
nonlocal current-field relation is shown to have non-trivial
consequences on the screening behavior of the normal metal. In the
ballistic case, we found the clean-limit theory to be restricted
further by $d < (\lambda^2(T)l)^{1/3}=\lambda_{\text{eff}}$,
$\lambda_{\text{eff}}$ giving the penetration depth for the non-local
current-field relation. If $\lambda_{\text{eff}}>d$, the screening
takes place on the geometric length scale $d$, leading to a saturation
at the screening fraction of $3/4$ at low temperatures. If
$\lambda_{\text{eff}}<d$, the finite (even though large) mean free
path strongly enhances the screening. Thus for typical samples with
$\lambda_N\ll d$ even a mean free path $l\gg d$ cannot be neglected,
i.e., the clean-limit behavior is practically unobservable. At large
temperatures $T\gg T_A$, a finite impurity concentration reduces the
range of the linear-response kernel to $\xi_T$, again enhancing the
screening. Furthermore, the screening fraction may serve to
distinguish between samples with bulk impurities rather than a rough
boundary, since a nonzero screening fraction at large temperatures is
{\em only} due to bulk impurity scattering. In dirty systems, where
the zeroth-order Green's function is well described by the Usadel
approximation, the current-field relation can still be non-local. We
have shown that the applicability of the local current-field relation
is restricted to $l^2\ll \lambda_N d$ for $T\ll E_{\text{Th}}$ and
$l^3\ll\lambda_N^2\xi_D(T)$ for $T\gg E_{\text{Th}}$. This shows that
in the presence of magnetic fields some caution is needed in applying
the Usadel theory.

We would like to acknowledge discussions with G. Blatter, G. Sch\"on,
F.~K.~Wilhelm, and A.~D.~Zaikin. A.~L.~F. is grateful to Universit\"at
Karlsruhe for hospitality, C.~B. and W.~B. would like to thank ETH
Z\"urich for hospitality and were supported by the Deutsche
Forschungsgemeinschaft (grant No.~Br1424/2-1) and the German Israeli
Foundation (Contract G-464-247.07/95).

\end{multicols}

\begin{thebibliography}{99}
\vspace*{-1cm}
\bibitem{mota:82}{
 A.~C. Mota, D. Marek, and J.~C. Weber,
 Helv. Phys. Acta {\bf 55}, 647 (1982).
}
\bibitem{pobell:87}{
 Th. Bergmann, K.~H. Kuhl, B. Schr\"oder, M. Jutzler, and F. Pobell,
 J. Low Temp. Phys. {\bf 66}, 209 (1987).
}
\bibitem{mota:90}{
 P. Visani, A.~C. Mota, and A. Pollini,
 Phys. Rev. Lett. {\bf 65}, 1514 (1990).
}
\bibitem{mota:94}{
 A.~C. Mota, P. Visani, A. Pollini, and K. Aupke,
 Physica B {\bf 197}, 95 (1994).
}
\bibitem{degennes}{
 Orsay Group on Superconductivity,
 in {\em Quantum Fluids}, ed. D. Brewer (North-Holland, Amsterdam, 1966).
}
\bibitem{zaikin}{
 A.~D. Zaikin,
 Solid State Commun. {\bf 41}, 533 (1982).
}
\bibitem{higashitani}{
 S. Higashitani and K. Nagai,
 J. Phys. Soc. Jpn. {\bf 64}, 549 (1995).
}
\bibitem{belzig}{
 W. Belzig, C. Bruder, and G. Sch{\"o}n, 
 Phys. Rev. B {\bf 53}, 5727 (1996).
}
\bibitem{fauchere}{
A.~L. Fauch\`ere and G. Blatter, Phys. Rev. B {\bf 56}, 14102 (1997). 
}
\bibitem{oda:80}{
 Y. Oda and H. Nagano,
 Solid State Commun. {\bf 35}, 631 (1980).
}
\bibitem{mota:89}{
 A.~C. Mota, P. Visani, and A. Pollini,
 J. Low Temp. Phys. {\bf 76}, 465 (1989).
}
\bibitem{oda:95}{
 H. Onoe, A. Sumuiyama, M. Nakagawa, and Y. Oda,
 J. Phys. Soc. Jpn. {\bf 64}, 2138 (1995).
}
\bibitem{london}{
F. and H. London, Proc. Phys. Soc., {\bf A149}, 71 (1938).
}
\bibitem{pippard}{
 A.~B. Pippard,
 Proc. Roy. Soc. {\bf 216}, 547 (1953).
}
\bibitem{eilenberger}{
 G. Eilenberger,
 Z. Phys. {\bf 214}, 195 (1968).
}
\bibitem{larkin}{
 A.~I. Larkin and Yu.~N. Ovchinnikov, Zh. Eksp.
 Teor. Fiz. {\bf 55} 2262 (1968) (Sov. Phys. JETP {\bf 26}, 1200
 (1968)).
}
\bibitem{rainersauls} For a comprehensive review of the quasiclassical
formalism applied to unconventional superconductors, see the
forthcoming book {\it Quasiclassical Methods in Superconductivity and
Superfluidity}, edited by D. Rainer and J.~A. Sauls (Springer,
Heidelberg, 1998).
\bibitem{herath}{
 J. Herath and D. Rainer,
 Physica C {\bf 161}, 209 (1989).
}
\bibitem{usadel}{
 K.~D. Usadel, 
 Phys. Rev. Lett. {\bf 25}, 507 (1970).
}

\bibitem{schopohl}{
 N. Schopohl and K. Maki,
 Phys. Rev. B {\bf 52}, 490 (1995), N. Schopohl, cond-mat/9804064.
}
\bibitem{eschrig}
  M. Eschrig, PhD thesis, Universit{\"a}t Bayreuth 1997 (unpublished);
  cond-mat/9804330 to be published in Ref.~\onlinecite{rainersauls}.
\bibitem{zaitsev}{
 A.~V. Zaitsev,
 Zh. Eksp. Teor. Fiz {\bf 86}, 1742 (1984)
 [Sov. Phys. JETP {\bf 59}, 1015 (1984)].
}
\bibitem{gorkov}{
 A.~A. Abrikosov, L.~P. Gorkov, and I.~E. Dzyaloshinski,
 {\em Methods of Quantum Field Theory in Statistical Physics},
 Dover (1975)
}
\bibitem{narikiyo:89}{
 O. Narikiyo and H. Fukuyama,
 J. Phys. Soc. Jpn. {\bf 58}, 4557 (1989).
}
\end{thebibliography}
\end{document}